\documentclass{llncs}
\usepackage{makeidx}
\usepackage{amssymb}
\usepackage{verbatimbox}
\usepackage{multirow}
\usepackage{scalerel}
\usepackage{mathtools}
\usepackage{circuitikz}
\usepackage{multicol}
\usepackage{amsmath, amssymb}% don't use amsthm, llncs.sty defines theorem macros
\usepackage{algpseudocode}
\usepackage{tikz}
\usepackage{cite}
\usetikzlibrary{arrows}
\usepackage{array, stackengine}
\newcommand{\extline}{$\scriptsize$-$\normalsize$\!}
\newcommand{\lextlineend}{$\scriptsize$\lhd\!$\normalsize$}
\newcommand{\rextlineend}{$\scriptsize\rule{.1ex}{0ex}$\rhd$\normalsize$}
\newcommand{\approxtext}[1]{\ensuremath{\stackrel{\text{#1}}{\equiv}}}
\newcolumntype{L}[1]{>{\raggedright\let\newline\\\arraybackslash\hspace{0pt}}m{#1}}
\newcolumntype{C}[1]{>{\centering\let\newline\\\arraybackslash\hspace{0pt}}m{#1}}
\newcolumntype{R}[1]{>{\raggedleft\let\newline\\\arraybackslash\hspace{0pt}}m{#1}}
\newcommand{\vpreceq}{\rotatebox[origin=c]{-90}{$\preceq$}}

\newcounter{index}

\usepackage[textsize=footnotesize,color=green!40]{todonotes}

\newcommand\blfootnote[1]{%
	\begingroup
	\renewcommand\thefootnote{}\footnote{#1}%
	\addtocounter{footnote}{-1}%
	\endgroup
}

\newcommand\extlines[1]{%
  \setcounter{index}{0}%
  \whiledo {\value{index}< #1}
  {\addtocounter{index}{1}\extline}
}

\newcommand\rextlinearrow[2]{$
  \setbox0\hbox{$\extlines{#2}\rextlineend$}%
  \tiny$%
  \!\!\!\!\begin{array}{c}%
  \mathrm{#1}\\%
  \usebox0%
  \end{array}%
  $\normalsize$\!\!%
}

\newcommand\lextlinearrow[2]{$
  \setbox0\hbox{$\lextlineend\extlines{#2}$}%
  \tiny%
  $%
  \!\!\!\!\begin{array}{c}%
  \mathrm{#1}\\%
  \usebox0%
  \end{array}%
  $\normalsize$\!\!%
}

\renewcommand\lextlinearrow[2]{%
  \setbox0\hbox{$\lextlineend\extlines{#2}$}%
  \shortstack{$\mathrm{#1}$\\\addvbuffer[-.7ex -.3ex]{\usebox0}}%
}

\renewcommand\rextlinearrow[2]{%
  \setbox0\hbox{$\extlines{#2}\rextlineend$}%
  \shortstack{$\mathrm{#1}$\\\addvbuffer[-.7ex -.3ex]{\usebox0}}%
}

\begin{document}

\frontmatter         

\pagestyle{headings}  
\mainmatter 

\title{How to Simulate It in Isabelle: Towards Formal Proof for Secure Multi-Party Computation}
\titlerunning{How to Simulate It in Isabelle: Towards Formal Proof for Secure Multi-Party Computation}  % abbreviated 
\author{David Butler\inst{1} \and David Aspinall\inst{1} and
Adri\`a Gasc\'on\inst{2}}

\institute{The Alan Turing Institute and University of Edinburgh
\and The Alan Turing Institute and University of Warwick}
\maketitle             

\begin{abstract}
  In cryptography, secure Multi-Party Computation (MPC) protocols
  allow participants to compute a function jointly while keeping their
  inputs private.  Recent breakthroughs are bringing MPC into
  practice, solving fundamental
  challenges for secure distributed computation.  Just as with classic protocols for encryption and key
  exchange, precise guarantees are needed for MPC designs and
  implementations; any flaw will give attackers a chance to break
  privacy or correctness.
  In this paper we present the first (as far as we know) formalisation
  of some MPC security proofs.  These proofs provide probabilistic
  guarantees in the computational model of security, but have a
  different character to machine proofs and proof tools implemented so far ---
  MPC proofs use a \emph{simulation} approach, in which
  security is established by showing indistinguishability between execution traces in the actual protocol execution and an ideal world where security is guaranteed by definition.  We show that existing machinery for
  reasoning about probabilistic programs adapted to this
  setting, paving the way to precisely check a new class of
  cryptography arguments. We implement our proofs using the CryptHOL framework inside Isabelle/HOL.

\end{abstract}

\section{Introduction}

Correctness \blfootnote{This work was supported by The Alan Turing Institute under the EPSRC grant EP/N510129/1.} guarantees are essential for cryptographic protocols and it is
an area where formalisation continues to have impact.
Older work was restricted to the \emph{symbolic (Dolev-Yao) model}~\cite{Dolev-Yao}, where cryptographic primitives are modelled as abstract operations and assumed to be unbreakable.  The symbolic model provides a baseline for correctness but modern cryptography is based on the more realistic \emph{computational model}~\cite{DBLP:journals/joc/AbadiR07}.  Adversaries are now allowed to break primitives, but are assumed to have limited computational power --- typically, polynomial time in a security parameter $n$, such as a key size.
Proofs in the computational model provide probabilistic guarantees: an adversary can break a security property only with negligible probability, i.e. probability bounded by a negligible function $\mu(n)$.
There are two main proof styles, the \emph{game-based} approach~\cite{Game} and the \emph{simulation-based} approach sometimes called the real/ideal world paradigm~\cite{Simulate}.

The simulation-based approach is a general proof technique especially useful for arguing about security of Multi-Party Computation (MPC) protocols.
MPC is an area of cryptography
concerned with enabling multiple parties to 
jointly evaluate a public function on their private inputs,
without disclosing unnecessary information (that is, 
without leaking any information about their respective inputs 
that cannot be deduced from their sizes or the result of the computation).
Several generic techniques can be used for that goal including Yao's garbled circuits~\cite{Yao86, Yaoproof},  
the GMW protocol~\cite{GMW}, and other protocols based on secret-sharing~\cite{Sharemind, spdz}.
These differ in whether they are designed for an arbitrary or fixed number of parties,
how the computed function is represented (e.g, Boolean vs. arithmetic circuits),
which functions can be represented (e.g, bounded-degree polynomials vs. arbitrary polynomials),
as well trade-offs regarding communication, computation requirements,  and
security guarantees.

In the last decade, groundbreaking developments have brought MPC closer to practice.
Efficient implementations of the protocols listed above
are available~\cite{mascot, oblivc, aby, oblivm}, and
we are now seeing the beginning of general solutions to fundamental security
challenges of distributed computation.
Security in these settings is proved by establishing a simulation
between the \emph{real world}, where the protocol plays out, and
an \emph{ideal world}, which is taken as the definition of security.
This formalises the intuition that
a protocol is secure if it can be simulated in an ideal environment
in which there is no data
leakage by definition.

A central protocol in MPC is Oblivious Transfer (OT), which allows a \emph{sender} to provide several values and a \emph{receiver} to choose some of them to receive, without learning the others, and without the sender learning which has been chosen.  In this paper we build up to a security proof of the Naor-Pinkas OT~\cite{N-P}, a practically important 1-out-of-2 oblivious transfer protocol (the receiver chooses one out of two messages).  This can be used as a foundation for more general MPC, as secure evaluation of arbitrary circuits can be based on OT~\cite{GMW}.

\paragraph{Contribution.}  As far as we know, this is the first
formalisation of MPC proofs in a theorem prover.  Our contributions are as follows.
\begin{itemize}
\item Starting from the notion of computational indistinguishablity,
  we formalise the simulation technique following the general form given
  by Lindell~\cite{Simulate}.
\item Lindell's method spells out a process but leaves details
  of reasoning to informal arguments in the cryptographer's mind;
  to make this fully rigorous, we use probabilistic programs to encode
  \emph{views} of the real and ideal worlds which can be successively
  refined to establish equivalence.  This is a general method which
  can be followed for other protocols and in other systems; it corresponds to \emph{hybrid arguments} often used in cryptography.
\item As examples of the method, we show information-theoretic
  security for a two-party secure multiplication protocol that uses a
  trusted initialiser, and a proof of security in the semi-honest
  model of the Naor-Pinkas OT protocol. The latter involves a reduction to the DDH assumption (a computational hardness assumption).
\item Finally, we demonstrate how a formalisation of security of a 1-out-of-2 OT can be extended to formalising the security of an AND gate.

\end{itemize}
We build on Andreas Lochbihler's
recent \emph{CryptHOL} framework~\cite{Andreas}, which provides tools for
encoding probabilistic programs using a shallow embedding inside Isabelle/HOL.
Lochbihler has used his framework for game-based cryptographic proofs, along similar
lines to proofs constructed in other theorem provers~\cite{CertiCrypt,FCF} and
dedicated tools such as EasyCrypt~\cite{EasyCrypt}. 

\paragraph{Outline.} In Sect. \ref{framework} we give an overview of the key parts of CryptHOL that we use and extend. Sect. \ref{comp-indist} shows how we define computational indistinguishability in Isabelle and Sect. \ref{sim_def} shows how it is used to define simulation-based security. In Sect. \ref{sim_game} we demonstrate how we use a probabilistic programming framework to do proofs in the simulation-based setting. Sect. \ref{Secure_mult} gives the proof of security of a secure multiplication protocol as a warm up and Sect.~\ref{N-P} shows the proof of security of the Naor-Pinkas OT protocol.  
In Sect. \ref{AND} we show how an OT protocol can be used to securely compute an AND gate, paving the way towards generalised protocols. Our formalisation is available online \cite{formalisation}.

\section{CryptHOL and Extensions} \label{framework}

CryptHOL is a probabilistic programming framework based around \emph{subprobability mass functions} (spmfs). An spmf encodes a discrete (sub) probability distribution. More precisely,  an spmf is a real valued function on a finite domain that is non negative and sums to at most one.  Such functions have type $\alpha \; \mathit{spmf}$ for a domain which is a set of elements 
of type $\alpha$.  We use the notation from~\cite{Andreas} and let $p!x$ denote the subprobability mass assigned by the spmf $p$ to the event $x$.  
The weight of an spmf is given by $||p|| = \sum_{y} p!y$ where the sum is taken over all elementary events of the corresponding type; this is the total mass of probability assigned by the spmf $p$. If $||p|| = 1$ we say $p$ is \emph{lossless}. Another important function used in our proofs is \emph{scale}. The expression $\mathit{scale} \; r \; p$ scales, by $r$, the subprobability mass of $p$. That is, we have $\mathit{scale} \; r \; p!x = r.(p!x)$ for $0 \leq r \leq \frac{1}{||p||}$.

Probabilistic programs can be encoded as sequences of functions that compute over values drawn from spmfs.  The type $\alpha \; \mathit{spmf}$ is used to instantiate the polymorphic monad operations $\mathit{return_{spmf}} :: \alpha \Rightarrow \alpha \; \mathit{spmf}$ and $\mathit{bind_{spmf}} :: \alpha \; \mathit{spmf} \Rightarrow (\alpha \Rightarrow \beta \; \mathit{spmf}) \Rightarrow \beta \; \mathit{spmf}$.

This gives a shallow embedding for probabilistic programs which we use to define simulations and views, exploiting the monadic do notation.  As usual, $do \; \{ x \leftarrow p; f \}$ stands for $\mathit{bind_{spmf}} \; p \; (\lambda x. \; do \; f)$. 

We note that $\mathit{bind_{spmf}}$ is commutative and constant elements cancel. In particular if $p$ is a lossless spmf, then
\begin{equation}\label{bind}
    \mathit{bind_{spmf}} \; p \; (\lambda \_.  \; q) = q.
\end{equation}

Equation \ref{bind} can be shown using the lemma \emph{bind\_spmf\_const}, 

\begin{equation}\label{bind_spmf_const}
 \mathit{bind_{spmf}} \; p \; (\lambda x. \; q) = \mathit{scale_{spmf}} \; (\mathit{weight_{spmf}} \; p) \; q \end{equation}
  and the fact $\mathit{sample\_uniform}$ is lossless and thus has weight equal to one. In Equation \ref{bind_spmf_const}, $\mathit{weight_{spmf}} \; p$ is $||p||$ described above.
  
The monad operations give rise to the functorial structure, $\mathit{map_{spmf}} :: (\alpha \Rightarrow \beta) \Rightarrow \alpha \; \mathit{spmf} \Rightarrow \beta \; \mathit{spmf}$. 

\begin{equation}\label{map_spmf_def}
\mathit{map_{spmf}} \; f \; p = \mathit{bind_{spmf}} \; p \; (\lambda x. \; \mathit{return_{spmf}} (f \; x)) \end{equation}

CryptHOL provides an operation, $\mathit{sample\_uniform}  ::  nat \Rightarrow nat \; \mathit{spmf}$ where $\mathit{sample\_uniform} \; n = \mathit{spmf\_of\_set} \; \{..<n\}$,
the lossless spmf which distributes probability uniformly to a set of
$n$ elements.  Of particular importance in cryptography is the uniform distribution
$\mathit{coin\_spmf} = \mathit{spmf\_of\_set} \; \{\mathsf{True},\mathsf{False}\}$. Sampling
from this corresponds to a coin flip. 

We also utilise the function $\mathit{assert\_spmf} ::  bool \Rightarrow \mathit{unit} \; \mathit{spmf}$ which takes a predicate and only allows the computation to continue if the predicate holds. If it does not hold the current computation is aborted. It also allows the proof engine to pick up on the assertion made.

One way we extend the work of CryptHOL is by adding one time pad lemmas needed in our proofs of security. We prove a general statement given in Lemma \ref{lemma:permute} and instantiate it prove the one time pads we require.

\begin{lemma}\label{lemma:permute}
  Let $f$ be injective and surjective on $\{..<q\}$.   
  Then we have
  $$\mathit{map_{spmf}} \; f \; (\mathit{sample\_uniform} \; q) = \mathit{sample\_uniform} \;   q.$$
\end{lemma}
\proof
By definition, $\mathit{sample\_uniform} \; q = \mathit{spmf\_of\_set} \; \{..<q\}$.
Then $\mathit{map_{spmf}} \; f  \\\; (\mathit{spmf\_of\_set} \; \{..<q\}) = \mathit{spmf\_of\_set} (f \; ^\backprime \; \{..<q\})$ follows by simplification and the injective assumption
(the infix $^\backprime$ is the image operator). Simplification uses the lemma \textit{map\_spmf\_of\_set\_inj\_on}:
$$\mathit{inj\_on} \; f \; A \implies\mathit{map_{spmf}} \; (\mathit{spmf\_of\_set} \; A) = \mathit{spmf\_of\_set} \; (f \; ^\backprime \; A).$$

We then have  $\mathit{map_{spmf}} \; f \; (\mathit{spmf\_of\_set} \; \{..<q\}) = \mathit{spmf\_of\_set} (\{..<q\})$ by using the surjectivity assumption. The lemma then follows from the definition of $\mathit{sample\_uniform}$.  
\qed

We note a weaker assumption, namely $f \: ^\backprime \: \{.. < q\} \subseteq \{.. < q\}$ can be used instead of the surjectivity assumption. To complete the proof with this assumption we use the $\mathit{endo\_inj\_surj}$ rule which states
$$\mathit{finite} \; A \implies f \; ^\backprime \; A \subseteq A \implies \mathit{inj\_on} \; f \; A \implies f \; ^\backprime \; A = A.$$

For the maps we use we prove injectivity and show surjectiveity using this.

\begin{lemma}[Transformations on uniform distributions]\label{lemmas}
  \begin{enumerate}
  \item $\mathit{map_{spmf}} \; (\lambda b. \; (y - b) \; mod \; q) \; (\mathit{sample\_uniform} \; q) = \mathit{sample\_uniform} \; q$.
   \label{transform-sub}
  \item $\mathit{map_{spmf}} \; (\lambda b. \; (y+b) \; mod \; q) \; (\mathit{sample\_uniform} \; q)  =  \mathit{sample\_uniform} \; q$.
  \item $\mathit{map_{spmf}} \; (\lambda b. \; (y +x.b) \; mod \; q) \;    (\mathit{sample\_uniform} \; q)  = \mathit{sample\_uniform} \; q$.
  \end{enumerate}
\end{lemma}
\proof These follow with the help of Lemma~\ref{lemma:permute}.
Case 3 holds only under the additional assumption that $x$ and $q$ are
coprime. This will always be the case in the applications we consider
as $x \in \mathbb{Z}_q$ and $q$ is a prime. \qed

\section{Computational Indistinguishability in Isabelle}\label{comp-indist}

We introduce the notion of computational indistinguishability as the definitions of security we give in Section \ref{sim_def} rely on it. We use the definition from \cite{Simulate}.

\begin{definition}\label{def_comp}
A probability ensemble $X = \{X(a,n)\}$ is a sequence of random variables indexed by $a \in \{0,1\}^\ast$ and $n \in \mathbb{N}$. Two ensembles $X$ and $Y$ are said to be computationally indistinguishable, written $X \approxtext{c} Y$, if for every non-uniform polynomial-time algorithm $D$ there exists a negligible function\footnote{A negligible function is a function $\epsilon \; :: \; \mathbb{N} \rightarrow \mathbb{R}$ such that for all $c \in \mathbb{N}$ there exists $N_c \in \mathbb{N}$ such that for all $x > N_c$ we have $|\epsilon(x)| < \frac{1}{x^c}$} $\epsilon$ such that for every $a$ and every $n \in \mathbb{N}$,

$$|Pr[D(X(a,n)) = 1] - Pr[D(Y(a,n)) = 1] | \leq \epsilon(n)$$
\end{definition}

The original definition restricts $a \in \{0,1\}^\ast$, but we generalise this
to an arbitrary first-order type, $\alpha$.  We model a probability ensemble as having some input of of this type, and a natural
number security size parameter. The space of events considered depends
on the \emph{view}; also of arbitrary first-order type, $\nu$.

$$\mathit{type\_synonym} \; (\alpha, \nu) \; \mathit{ensemble} = \alpha \Rightarrow nat \Rightarrow \nu\; \mathit{spmf}$$

We do not formalise a notion of polynomial-time programs in Isabelle as we do not need it
to capture the following proofs.
In principle this could be done with a deep embedding of a programming language, its semantic denotation function and a complexity measure. 
Instead, we will assume a family of constants giving us the set of all
polynomial-time distinguishers for every type $\nu$, indexed by a
size parameter. 

A polynomial-time distinguisher ``characterises'' an arbitrary spmf.  

$$\mathit{consts} \; \mathit{polydist} \; :: \; nat \Rightarrow (\nu \; \mathit{spmf} \Rightarrow \mathit{bool \; spmf}) \; \mathit{set}$$

Now we can formalise Definition \ref{def_comp} directly as:
\begin{algorithmic}
\State $\mathit{comp\_indist} \;:: \; (\alpha, \nu) \; \mathit{ensemble} \Rightarrow (\alpha, \nu) \; \mathit{ensemble} \Rightarrow \mathit{bool} $
\State $\mathit{where} \; \mathit{comp\_indist} \; X \; Y \equiv$
\State $\quad \forall (D \; :: \; \nu \; \mathit{spmf} \Rightarrow \mathit{bool \; spmf}).$
\State $\quad \qquad \exists \; (\epsilon \; :: \; \mathit{nat} \Rightarrow \mathit{real}). \; \mathit{negligible} \; \epsilon \; \wedge$
\State $\quad \qquad \qquad (\forall \; (a \; :: \; \alpha) \; (n \; :: \; \mathit{nat}). $
\State $\quad \qquad \qquad \qquad (D \; \in \; \mathit{polydist} \; n) \longrightarrow $
\State $\quad \qquad \qquad \qquad \quad |\mathit{spmf} \; (D \; (X \; a \; n)) \; \mathit{True}  -  \mathit{spmf} \; (D \; (Y \; a \; n)) \; \mathit{True}| \leq \epsilon \; n))$
\end{algorithmic} 

\section{Semi-Honest Security and Simulation-Based Proofs} \label{sim_def}

In this section we first define security in the semi-honest adversary model using the simulation-based approach. We then  show how we use a probabilistic programming framework to formally prove security. 

A protocol is an algorithm that describes the interaction between parties and can be modelled as a set of probabilistic programs. A two party protocol $\pi$ computes a map from pairs of inputs to pairs of outputs. This map is called the protocol's \emph{functionality} as it represents the specification of what the protocol should achieve. It can be formalised as a pair of (potentially probabilistic) functions
$$f_1 : input_1 \times input_2 \longrightarrow output_1$$
$$f_2 : input_1 \times input_2 \longrightarrow output_2$$

\noindent which represent each party's output independently. The composed pairing is the functionality, $f$, of type  

$$f : input_1 \times input_2 \longrightarrow output_1 \times output_2$$

\noindent where $f = (f_1, f_2)$. That is, given inputs $(x, y)$ the functionality outputs $(f_1(x,y), \\f_2(x,y))$. This indicates that party one gets $f_1(x,y)$ and party two gets $f_2(x,y)$ as output. In general the types of inputs and outputs can be arbitrary. For our instantiation we use concrete types depending on the functionality concerned.

For the initial example secure multiplication protocol we consider in Section \ref{Secure_mult} we have the probabilistic functionality  $f(x,y) = (s_1, s_2) $ where $s_1 + s_2 = x.y$. Each party obtains an additive share of the multiplication. The protocol is run using a publicly known field $\mathbb{Z}_q$ where $q$ is a prime number dependent on the security parameter. To ensure neither of the outputs alone reveal the value of $x.y$, we uniformly sample one of the outputs in the functionality

\begin{equation}\label{sec_mult_funct}
   f(x,y) = (s_1, x.y - s_1), s_1 \xleftarrow[]{\$} \mathbb{Z}_q
\end{equation}

The notation $s_1 \xleftarrow[]{\$} \mathbb{Z}_q$ means we sample $s_1$ uniformly from $\mathbb{Z}_q$. The Isabelle definition of the functionality is given below. It makes use of the do notation:

	\begin{center}
		\begin{tabular}{l}
			$f \; x \; y = do \; \{$ \\
			$\qquad s_1  \leftarrow \mathit{sample\_uniform} \; q;$ \\
			$\qquad \mathit{return_{spmf}} (s_1, x.y - s_1) \}$ \\
		\end{tabular} 
	\end{center} 

This functionality is easy to compute if one does not consider security; the parties can share their inputs and compute it. But with the security requirement that neither party learns anything about the others' input the problem becomes harder. We will give a protocol that securely computes this functionality later. We first introduce the notions used to define security. Security is based on \emph{views} which capture the information known by each party. We follow the definitions given by Lindell in \cite{Simulate} to define security in the semi-honest model.

\begin{definition}
Let $\pi$ be a two party protocol with inputs $(x,y)$  and with security parameter $n$.
\begin{itemize}
    
\item  The real view of the $i^{th}$ party (here $i \in \{1,2\}$) is denoted by $$view^{\pi}_i (x,y,n) = (w,r^i,m^i_1,...,m^i_t)$$ where $w \in \{x,y\}$ and is dependent on which view we are considering, $r^i$ accumulates random values generated by the party during the execution of the protocol, and the $m_j^{i}$ are the messages received by the party.  
 
\item Denote the output of the $i^{th}$ party, $output^{\pi}_i(x,y,n)$, and the joint output as $$output^{\pi}(x,y,n) = (output^{\pi}_1(x,y,n), output^{\pi}_2(x,y,n)).$$
 \end{itemize}   
\end{definition}

\begin{definition}  \label{non_det}
A protocol $\pi$ is said to securely compute $f$ in the presence of a semi-honest adversary if there exist probabilistic polynomial time algorithms (simulators) $S_1,S_2$ such that  $$\{S_1(1^n,x,f_1(x,y)),f(x,y)\}   \approxtext{c} \{view^{\pi}_1(x,y,n), output^{\pi} (x,y,n)\}$$
$$\{S_2(1^n,y,f_2(x,y)),f(x,y)\}   \approxtext{c} \{view^{\pi}_2(x,y,n), output^{\pi} (x,y,n)\}.$$ 
\end{definition} 

A semi-honest adversary is one that follows the protocol description. The simulator is given a unary encoding of the security parameter.

This definition formalises the idea that a protocol is secure if whatever can be computed by a party can also be computed from only the input and output of the party meaning that nothing extra is learned from the protocol.

For the secure multiplication protocol and the receiver's security in the Naor-Pinkas OT we prove security in an information theoretic sense. This means even computationally unbounded adversaries cannot gain extra information from the protocol. This is shown by proving the two sets of distributions above are equal. Information theoretic security is a stronger notion of security than computational indistinguishability and Isabelle proves the former implies the latter with ease.

A functionality is deterministic if given inputs always produce the same output. For a deterministic protocol it is shown in \cite{Simulate} that the above definition can be relaxed. We require correctness and
\begin{equation}\label{non-det1}
\{S_1(1^n,x,f_1(x,y))\}   \approxtext{c} \{view^{\pi}_1(x,y,n)\} 
\end{equation}
\begin{equation}\label{non-det2}
\{S_2(1^n,y,f_2(x,y))\}   \approxtext{c} \{view^{\pi}_2(x,y,n)\}
\end{equation}
For a protocol to be correct we require that for all $x,y$ and $n$ there exists a negligible function $\mu$ such that 
$$Pr[output^{\pi} (x,y,n) \not = f(x,y)] \leq \mu(n).$$

The Naor-Pinkas OT protocol,  and the OT we use in the AND gate protocol given later, are both deterministic. The secure multiplication protocol however is not. For the deterministic cases we will focus on the more interesting property, showing the views are equal. As such when we refer to a deterministic protocol as being secure we explicitly show Equations \ref{non-det1} and \ref{non-det2} and assume correctness. For the non-deterministic secure multiplication protocol we must show exactly the property given in Definition \ref{non_det}.

\subsection{Probabilistic Programming used for Simulation-Based Proofs }\label{sim_game}
\begin{figure}
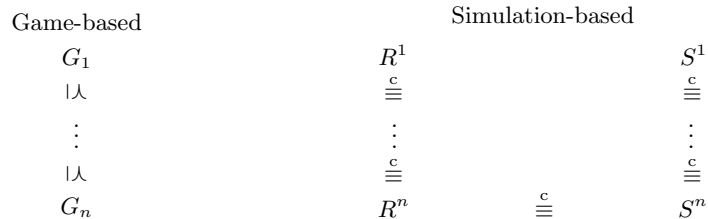
 
\centering
\begin{tabular}{C{3cm} C{2cm} C{5cm}}
  Game-based & & Simulation-based \\[1.3ex]
  $G_1$    & & $R^1$ \hspace{3.40cm} $S^1$ \\
  \vpreceq & &  $\approxtext{c}$     \hspace{3.5cm} $\approxtext{c}$     \\
  \vdots & &  \vdots    \hspace{3.75cm} \vdots     \\
  \vpreceq & &  $\approxtext{c}$  \hspace{3.5cm} $\approxtext{c}$     \\
  $G_n$    & & $R^n$ \hspace{1.55cm}$\approxtext{c}$\hspace{1.55cm}  $S^n$ \\
\end{tabular}
\caption{A comparison between the game-based and simulation-based approaches. The game-based approach uses reductions (denoted $\preceq$) whereas in the simulation approach we show computational indistinguishability between probabilistic programs.} \label{game-sim}
\end{figure}

CryptHOL provides a strong foundation from which to manipulate and show equivalence between probabilistic programs. So far it has only been used to prove security in the game-based setting. The game-based definitions of security use a game played between an adversary and a benign challenger. The players are modelled as probabilistic programs and communicate with each other. The definition of security is tied to some event which is defined as the output of the security game. In general, proofs describe a reduction of a sequence of games (probabilistic programs) that end in a game where it can be shown the adversary has the same advantage of winning over the challenger as it would have against a problem assumed to be hard. The games in the sequence are then shown to be equivalent. This is shown on the left hand side of Fig. \ref{game-sim}.
We use a probabilistic programming framework to construct simulation-based proofs.
Our method of proof models the simulator and the real view of the protocol as probabilistic programs. In the right hand side of Fig. \ref{game-sim} we start with the real view of the protocol, $R^1$, and the simulator, $S^1$. We define a series of intermediate probabilistic programs ($R^i, S^i$) which we show to be computationally indistinguishable (or equal in the case of information theoretic security) --- this is referred to as the \emph{hybrid argument} in cryptography. This sequence ends in $R^n$ and $S^n$ which we show to be computationally indistinguishable (or equal). We have shown the diagram for the simulation-based approach in Fig. \ref{game-sim} is transitive.

\begin{lemma} Let $X$, $Y$ and $Z$ be probability ensembles then we have
$$[ X \approxtext{c} Y ; \; Y \approxtext{c} Z] \implies X \approxtext{c} Z. $$
\end{lemma}

For the non-deterministic secure multiplication protocol we will construct the protocol and functionality outputs in the real and simulated views, instead of constructing them separately and combining them to form the ensembles.

\section{Secure Multiplication Protocol} \label{Secure_mult}

We now present a protocol that computes the functionality in Equation \ref{sec_mult_funct}. The protocol requires some pre-generation of elements to be distributed to the parties. This is known in MPC as the preprocessing model~\cite{Beaver-triples}, where the parties run an offline phase to generate correlated random triples --- sometimes called Beaver triples --- that are used to perform fast secure multiplications in an online phase. For this task we assume a trusted initialiser that aids in the computation. 
\begin{figure}
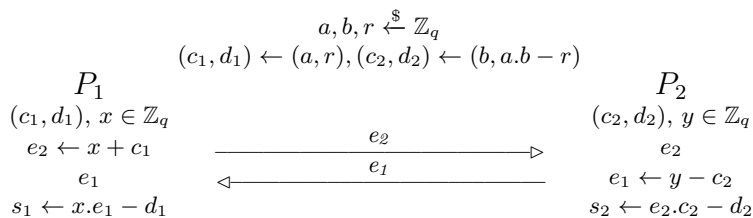
 
\begin{center} 
\begin{tabular}{ccc}
&$ a, b, r \xleftarrow[]{\$} \mathbb{Z}_q$ & \\& $(c_1, d_1) \leftarrow (a,r), (c_2,d_2) \leftarrow (b, a.b -r)$& \\
{\large $P_1$} & & {\large $P_2$}\\
$(c_1, d_1)$, $x \in \mathbb{Z}_q$ & & $(c_2, d_2)$, $y \in \mathbb{Z}_q$  \\
$e_2 \leftarrow x + c_1$ & \rextlinearrow{\mathit{e_2}}{26} & $e_2$ \\
$e_1$ & \lextlinearrow{\mathit{e_1}}{26} & $e_1 \leftarrow y - c_2$ \\
$s_1 \leftarrow x.e_1 - d_1$ & & $s_2 \leftarrow e_2.c_2 - d_2$\\
\end{tabular}
\end{center}
\caption{A protocol for secure multiplication} 
\label{SM_PROT_S}
\end{figure}
We denote the assignment of variables by $a \leftarrow b$ and all operations are taken modulo $q$. The claim of security is: 

\begin{theorem}\label{sec_mult_thm}
The protocol in Fig. \ref{SM_PROT_S} securely computes the functionality given in Equation \ref{sec_mult_funct} in the semi-honest adversary model.
\end{theorem}

Intuitively, security results from the messages being sent in the protocol always being masked by some randomness. In the message party one sends, $e_2$, the input ($x$) is masked by the uniform sample, $c_1$. Likewise in the message party two sends, $e_1$, the input ($y$) is masked by the uniform sample, $c_2$.

\subsection{Formal Proof of Security}

The simulator and the real view of party one are defined in Isabelle as in Fig. \ref{S1R1}.  Recall that the simulator receives as input the input of the party it is simulating and the output of the functionality of the party it is simulating. 

\begin{figure}
\begin{center}
\begin{tabular}{l C{0.5cm} l}
$S_1 \; x \; s_1' = do \; \{$ && $R_1 \; x \; y = do \; \{$ \\
$\qquad c_1'  \leftarrow \mathit{sample\_uniform} \; q;$ && $\qquad ((c_1, d_1), (c_2, d_2)) \leftarrow  TI;$\\
$\qquad e_1' \leftarrow \mathit{sample\_uniform} \; q;$ && $\qquad \mathit{let} \; e_1 = (y - c_2)\; mod \; q;$\\
$\qquad \mathit{let} \; d_1' = (x.e_1' - s_1')\; mod \; q;$ && $\qquad \mathit{let} \; e_2 = (x + c_1)\; mod \; q;$\\
$\qquad \mathit{let} \; s_2' = (x.y - s_1') \; mod \; q;$ && $\qquad \mathit{let} \; s_1= (x.e_1 - d_1)\; mod \; q;$\\
$\qquad \mathit{return_{spmf}} (x, c_1', d_1', e_1', s_1', s_2') \}$ && $\qquad \mathit{let} \; s_2 = (e_2.c_2 - d_2)\; mod \; q;$\\
&& $\qquad \mathit{return_{spmf}} (x, c_1, d_1, e_1, s_1, s_2) \}$ \\ 
\end{tabular} 
\end{center} \caption{Probabilistic programs to output the real and simulated views for party one. Here, TI is the trusted initialiser.}\label{S1R1}
\end{figure}

Note that the simulator $S_1$ takes uses $y$ in the construction of the functionality output. This is allowed by the security definition as $\mathit{f(x,y)}$ depends on both inputs.  

To show information theoretic security we prove that the two probabilistic programs given in Fig \ref{S1R1} are equal - when $s_1'$, the input to the simulator is the first output of the functionality. This involves a series of small equality steps between intermediate probabilistic programs as shown in Fig \ref{game-sim}. In particular, in the series of intermediate programs we manipulate the real and simulated views. We note that $\mathit{c_1', e_1'}$ and $s_1'$ are random samples that are independent from each other, $x$ and $y$ and we have $s_2' = (x.y - s_1') \; mod \; q$ and $d_1' = (x.e_1' - s_1') \; mod \; q$. By showing these relationships, and only these relationships, hold for the real view too we show the two views are equal.

This gives us the first half of formal security which can be seen in Lemma \ref{P1}

\begin{lemma} \label{P1} For all inputs $x$ and $y$ we have, $S_1 \; x \; y = R_1 \; x \; y$.  This
  implies the definition of security we gave in Sect.~\ref{sim_def},
  $S_1 \; x \; y \approxtext{c} R_1 \; x \; y$.
\end{lemma}

The proof of security for party two is similar, where the relationships to consider are $s_2' = (x.y - s_1') \; mod \; q$ and $d_2' = (e_2'.c_2' - s_2') \; mod \; q$. Together, Lemmas \ref{P1} and \ref{P2} establish Theorem \ref{sec_mult_thm}.

\begin{lemma} \label{P2} For all inputs $x$ and $y$ we have, 
$S_2 \; x \; y = R_2 \; x \; y$. This implies the definition of security we gave in Sect.~\ref{sim_def}, $S_2 \; x \; y \approxtext{c} R_2 \; x \; y$.
\end{lemma}

\section{Naor-Pinkas Protocol} \label{N-P}

\begin{figure}
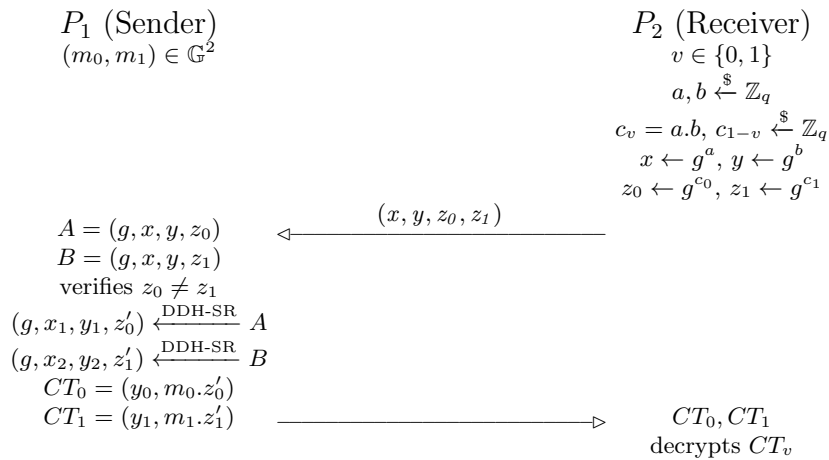

\begin{center}
\begin{tabular}{clc}
{\large $P_1$ (Sender)} & & {\large $P_2$ (Receiver) }\\
$(m_0, m_1) \in \mathbb{G}^2$ & & $v \in \{0,1\}$  \\
& & $a, b \xleftarrow{\$} \mathbb{Z}_q $ \\
& & $c_v = a.b$,  $c_{1-v} \xleftarrow{\$} \mathbb{Z}_q $ \\
& & $x \leftarrow g^a$, $y \leftarrow g^b$ \\
&& $z_0 \leftarrow g^{c_0}$, $z_1 \leftarrow g^{c_1}$ \\
$A = (g, x, y, z_0)$ & \lextlinearrow{\mathit{(x, y, z_0, z_1)}}{26} &\\
$B = (g, x, y, z_1)$ & & \\
verifies $z_0 \neq z_1$ && \\ 
$(g, x_1, y_1, z_0')\xleftarrow{\text{DDH-SR}} A$  && \\ $(g, x_2, y_2,  z_1') \xleftarrow{\text{DDH-SR}} B$ &&\\
$CT_0 = (y_0, m_0.z_0')$ && \\  $CT_1 = (y_1, m_1.z_1')$ & \rextlinearrow{}{26} & $CT_0, CT_1$ \\
&& decrypts $CT_v$
\end{tabular}
\end{center}
\caption{The Naor-Pinkas OT protocol} \label{N-P_prot}
\end{figure}

In the Naor-Pinkas OT protocol~\cite{N-P} we work with a cyclic group $\mathbb{G}$ of order $q$ where $q$ is a prime, for which the DDH assumption holds. The Decisional Diffie Hellman (DDH) assumption \cite{DBLP:journals/tit/DiffieH76} is a computational hardness assumption on cyclic groups. Informally, the assumption states that given $g^a$ and $g^b$, where $a$ and $b$ are uniform samples from $\mathbb{Z}_q$, the group element $g^{a.b}$ looks like a random element from $\mathbb{G}$. A triple of the form $(g^a, g^b, g^{a.b})$ is called a DDH triple. 
 In the protocol, given in Fig \ref{N-P_prot}, the Sender (party one) begins with input messages $(m_0, m_1) \in \mathbb{G}^2$ and the Receiver (party two) begins with $v \in \{0,1\}$, the choice bit. At the end of the protocol the receiver will know $m_v$ but will learn nothing about $m_{1-v}$ and the sender will not learn $v$.
 
 We prove information theoretic security in the semi-honest model for the receiver. Security for the sender is proven with a reduction to the DDH assumption. 
In particular, the receiver is only able to decrypt $m_v$ as the corresponding ciphertext is a valid ElGamal ciphertext, while $m_{1-v}$ is garbage. 

In the protocol description, given in Fig \ref{N-P_prot}, DDH-SR refers to a DDH \emph{random self reduction} operation which takes DDH triples to DDH triples and non DDH triples to non DDH triples. The reduction is defined as follows. Given an input tuple $(g, g^x, g^y, g^z)$, one picks $a,b$ uniformly from $\mathbb{Z}_q$ and outputs $(g, g^{(x+b)a}, g^ y, g^{(z+b.y)a})$. The role of the DDH random self reduction is to destroy any partial information in the message the Receiver sends to the Sender.

\begin{theorem}
The protocol defined in Fig. \ref{N-P_prot} securely computes a 1-out-of-2 OT in the semi-honest adversary model.
\end{theorem}

\subsection{The Formal Proof} \label{N-P proof}

We have a deterministic protocol and so do not include the overall functionality as part of the views. We must first consider the DDH-SR. In particular the two cases, when the input tuple is a DDH triple and when it is not. In both cases we simplify the operation that is performed. The simplified definitions are given in Fig \ref{simplified_DDH_SR} and the formal statements in Lemmas \ref{DDH_tuple} and \ref{DDH_non_tuple}:

\begin{figure} 
	\begin{center} 
		\begin{tabular}{l C{0.2cm} l}
			$DDH\_SR\_triple\; x \; y \; z = do \; \{$ && $\mathit{DDH\_SR\_non\_triple} \; x \; y \; z = do \; \{$ \\
			$\qquad x_1 \leftarrow \mathit{sample\_uniform} \; q;$ && $\qquad x_1, x_2 \leftarrow \mathit{sample\_uniform} \; q;$ \\
			$\qquad \mathit{return_{spmf}} (g, g^{x_1}, g^y,g^ {y.x_1 \; mod \; q}) \}$ && $\qquad \mathit{return_{spmf}} (g, g^{x_1}, g^y, g^{x_2})\}$  \\
		
		\end{tabular} 
	\end{center}\caption{The two simplified probabilistic programs for the DDH triples and non-triples.} \label{simplified_DDH_SR}
\end{figure} 

\begin{lemma} \label{DDH_tuple}
	For all $x,y,z$ such that $z = y.x \; mod \; q$  we have $$DDH\_SR \; x \; y \; z = DDH\_SR\_triple \; x \; y \; z.$$
\end{lemma}

\begin{lemma} \label{DDH_non_tuple}
	For all $x,y,z$ such that $z \not = y.x \; mod \; q$  we have $$DDH\_SR \; x \; y \; z = DDH\_SR\_non\_triple\; x \; y \; z.$$
\end{lemma}

\subsubsection{The Simulators and Views.}
 First we consider party two. In constructing the real and simulated views we use the assert function to ensure the condition given in the protocol in Fig \ref{N-P_prot}, $z_0 \not = z_1$, holds. This ensures that only one of $A$ and $B$ is a DDH triple; the other is not and hence the corresponding ciphertext $CT_0$ or $CT_1$ cannot be decrypted. The simulator may take as inputs $v \in \{0,1\}$ and $CT_v$ (although does not require it). We use $\otimes$ to denote multiplication in the group (as in Isabelle). The real view and simulator are shown below.

\begin{center}
\begin{tabular}{l  l}
$S_2 \; v = do \; \{$ & $R_2\; m_0 \; m_1 \; v = do \; \{$ \\
$\quad a, b \leftarrow \mathit{sample\_uniform} \; q;$ & $\quad a,b \leftarrow \mathit{sample\_uniform} \; q;$ \\
$\quad let \; c_v = a.b ;$ & $\quad let \; c_v = a.b;$  \\
$\quad c_v' \leftarrow \mathit{sample\_uniform} \; q;$ & $\quad c_v' \leftarrow \mathit{sample\_uniform} \; q ;$\\
$\quad \_ \leftarrow \mathit{assert\_spmf} (c_v'\not = b.a \; mod \; q) ;$ & $\quad \_ \leftarrow \mathit{assert\_spmf} (c_v'\not = b.a \; mod \; q) ;$\\
$\quad x_0 \leftarrow \mathit{sample\_uniform} \; q ;$ &  $\quad (g, x_0, y_0,z_0') \leftarrow \mathit{DDH\_SR} \; a \; b \; c_v;$\\
$\quad x_1 \leftarrow \mathit{sample\_uniform} \; q ;$ &  $\quad (g, x_1, y_1,z_1') \leftarrow \mathit{DDH\_SR} \; a \; b \; c_v';$\\
$\quad \mathit{return_{spmf}} (v, a, b, c_v', g ^ b, g ^ {x_1}, g ^ b, g^{x_2}) \}$& $\quad let \; e_0 = z_0' \otimes m_0; $\\
& $\quad let \; e_1 = z_1' \otimes m_1;$ \\
& $\quad \mathit{return_{spmf}} (v, a, b,c_v', y_0, e_0, y_1, e_1) \}$ \\
\end{tabular}
\end{center}

For party one, the simulator, $S_1$, takes in the two messages $(m_0, m_1)$ (again, it does not use them) and the Sender's output - which amounts to nothing. The simulator and real view are given below. We note there is no output for the sender from the functionality so the simulator only takes the messages as input.

\begin{center}
\begin{tabular}{l C{0.5cm} l}
$S_1 \; (m_0, m_1) = do \; \{$ &&  $R_1 \; (m_0, m_1) \; v = do \; \{$\\
$\quad a,b, c \leftarrow \mathit{sample\_uniform} \; q;$ && $\quad a, b, c_v' \leftarrow \mathit{sample\_uniform} \; q;$\\
$\quad \mathit{return_{spmf}} (g^a, g^b, g^{a.b}, g^ {c_1}) \}$ && $\quad let \; c_v = a.b;$ \\ 
 && $\qquad \mathit{return_{spmf}} ((m_0, m_1), (g^a, g^b, \mathit{ if \: v \: then} $\\
 && $ \qquad \quad \mathit{g^{c_v'} \: else \: g^{c_v}, \: if \: v \: then \: g^{c_v} \: else \: g^{c_v'}))} \}$ \\
\end{tabular}  %\caption{The probabilistic program outputting the simulated view for party one}
\end{center}

\subsubsection{Proof of Security for the Receiver.}

From the construction of the real view one can see the triple $(a, b, c_v)$ is a DDH triple and $(a, b, c_v')$ is not. Thus we are able to rewrite the real view using Lemmas \ref{DDH_tuple} and \ref{DDH_non_tuple}. 

The only components of the outputs of $R_2$ and $S_2$ which differ, up to unfolding of definitions are the encryptions. In the real view they are of the form $g^z \otimes m_i$ where $z$ is uniformly sampled and in the simulator they are of the form $g^z$. We utilise a lemma from CryptHOL which states that if $c \in \mathit{carrier} \; \mathbb{G}$ then:
\begin{multline*}
    \mathit{map_{spmf}} \; (\lambda x. \; g^x \otimes c) \; (\mathit{sample\_uniform} \; q)  \\= \mathit{map_{spmf}} \; (\lambda x. \; g^x) \; (\mathit{sample\_uniform} \; q)
\end{multline*}

This allows us to show our security result stated in Lemma \ref{R2_S2}.
\begin{lemma} \label{R2_S2} 
	For all inputs $m_0$, $m_1$ and $v$ we have, $S_2 \;  v = R_2 ; m_0 \; m_1 \; v$.  This
	implies the definition of security we gave in Sect.~\ref{sim_def},
	$S_2 \;  v \approxtext{c} R_2 ; m_0 \; m_1 \; v$.
\end{lemma}

\subsubsection{Proof of security for the Sender.}
For $v=0$, the proof is trivial as the simulator and real views are constructed in exactly the same way.

\begin{lemma} \label{Sender_v_0}
The case of $v=0$ for party one implies for all inputs $m_0$ and $m_1$, $$R_1\; m_0 \; m_1= S_1 \; m_0 \; m_1.$$
\end{lemma}
The proof for $v=1$ is equivalent to showing the distributions $(g ^ a, g^b, g^{a.b}, g ^ c)$ and $(g ^ a, g^b, g^c, g ^ {a.b})$ are computationally indistinguishable, when $a, b,c$ are uniformly sampled. Here we provide a high level view of the pencil and paper. 

To show security we provide a reduction to the DDH assumption, which implies the two distributions are computationally indistinguishable. In particular we show that if there exists a $D$ that can distinguish the above two 4-tuples then one can construct an adversary that breaks the DDH assumption.  We use the formalisation of the DDH assumption from \cite{Andreas}.

\begin{definition}
 The DDH advantage for a distinguisher $D$ is defined as

$$adv\_ddh(D)= Pr[D((g^a, g^b, g^{a.b})) = 1] - Pr[D((g^a, g^b, g^c))= 1]$$

where $a, b, c \xleftarrow{\$} \mathbb{Z}_q.$
\end{definition}

We show the reduction in two steps. First we show a reduction from $(g^a, g^b, g^c, g^{a.b})$ to $(g^a, g^b, g^c, g^d)$ and then from $(g^a, g^b, g^c, g^d)$ to $(g^a, g^b, g^{a.b}, g^c)$ where $a, b, c, d$ are all uniform samples. Consider the first reduction - we assumes $D$ can distinguish the tuples. 

\paragraph{DDH Adversary one (Adv1), inputs: $ D, (\alpha, \beta, \gamma)$ .}
\begin{itemize}
	\item The adversary samples $c \leftarrow \mathit{sample\_uniform} \; (order \; \mathcal{G})$.
	\item The provides D with the input $(\alpha, \beta, g^c, \gamma)$ and outputs whatever $D$ outputs. 
\end{itemize}

The second reduction (using $Adv2$) is analogous, 

Using both of these we can prove security by bounding the advantage an adversary has of distinguishing between the real and simulated views by the sum of two DDH advantages, which are assumed to be negligible.

\begin{lemma} 
\begin{multline*}
\mathit{spmf \; (\mathcal{A} \; D \; (R1 \; (m0,m1) \; v)) \; True - spmf \; (\mathcal{A} \; D \; (S1 \; (m_0,m_1))) \; True  \leq} \\ \mathit{ddh.adv \; (Adv1 \; D \; (m_0, m_1)) + ddh.adv \; (Adv2 \; D \; (m_0, m_1))} 
\end{multline*}
\end{lemma}

This along with showing information theoretic security (Lemma \ref{R2_S2}) for the receiver means we have shown the protocol to be secure in the semi-honest model.
 
\section{Towards Evaluating Arbitrary Functionalities} \label{AND}

Several MPC techniques allow for the secure joint evaluation
of {\em any} functionality represented as a Boolean circuit
or an arithmetic circuit.
At a high level, these protocols proceed by evaluating
the circuit gate by gate while always keeping a secret share of the partial evaluation.
In particular the GMW protocol relies on OT to securely evaluate AND gates

\begin{figure}
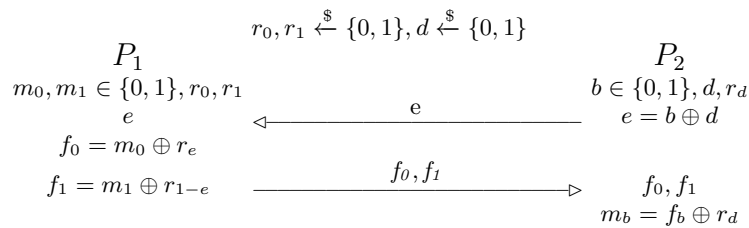
 
	\begin{center} 
		\begin{tabular}{clc}
			& $r_0, r_1 \xleftarrow{\$} \{0,1\}, d \xleftarrow{\$} \{0,1\}$ & \\
			{\large $P_1$} & & {\large $P_2$}\\
			$m_0, m_1 \in \{0,1\}, r_0, r_1$ && $b \in \{0,1\}, d, r_d$ \\
					$ e $ & \lextlinearrow{ e}{26} & $e = b \oplus d$ \\
			$f_0 = m_0 \oplus r_e$ && \\
	$f_1 = m_1 \oplus r_{1-e}$ & \rextlinearrow{ \mathit{f_0, f_1}}{26} & $f_0, f_1$ \\ 
			& & $m_b = f_b \oplus r_d$\\
		\end{tabular} \caption{Single bit OT}\label{a}
	\end{center}
\end{figure}

\begin{figure}
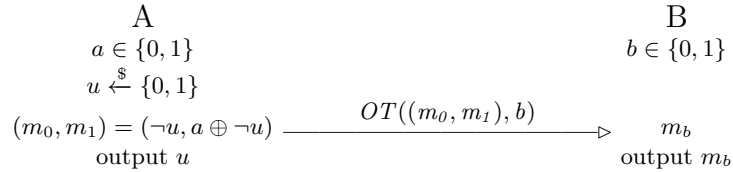
 
\begin{center} 
\begin{tabular}{clc}
{\large A} & & {\large B}\\
$a \in \{0,1\}$ & & $b \in \{0,1\}$ \\
$u \xleftarrow{\$} \{0,1\}$ & &  \\
$(m_0, m_1) = (\neg u, a \oplus \neg u)$ & $\rextlinearrow{\mathit{OT((m_0,m_1), b)}}{26}$ & $m_b$ \\
output $u$ & & output $m_b$\\
\end{tabular} \caption{A protocol to compute an AND gate}\label{b}
\end{center}
\end{figure}

In this section we use  a basic OT protocol (Fig \ref{a}) to construct a protocol to
compute the output of an AND gate. The OT protocol we use employs a trusted initialiser, like the secure multiplication protocol of Section \ref{Secure_mult}. The trusted initialiser pre-distributes correlated randomness to the parties so they can carry out the protocol. In particular $r_0$ and $r_1$ are uniformly sampled and given to party one, and $d$ is uniformly sampled and given to party two along with $r_d$. The AND gate protocol then uses OT, this is done in a similar way as in the GMW protocol.
The AND gate protocol we use here is taken from \cite{DBLP:conf/crypto/BennettBCS91}  and is described in Fig. \ref{b}. This demonstrates that OT can be used in powerful ways to construct protocols to compute fundamental functions securely.

Initially we show information theoretic security for the OT construction given in Fig \ref{a}. That is we construct simulators $S_1^{OT}$ and $S_2^{OT}$ such that for the appropriately defined views $R_1^{OT}$ and $R_2^{OT}$ the result in Lemma \ref{lem_OT} holds. To do this we define an appropriate XOR function ($\oplus$) on Booleans and prove a one time pad lemma on the XOR function.

\begin{lemma}\label{lem_OT}
	$R_1^{OT} \; m_0 \;  m_1 \; b = S_1^{OT} \; m_0 \; m_1$ and $R_2^{OT} \; m_0 \; m_1 \; b = S_2^{OT} \; b.$
\end{lemma}

\begin{figure}
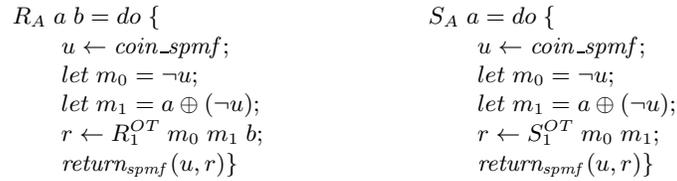

\begin{center}
\begin{tabular}{l C{2cm} l}
 $R_A \; a \; b  = do \; \{$ &  & $S_A \; a = do \; \{$ \\
$\qquad u \leftarrow \mathit{coin\_spmf};$ && $\qquad u \leftarrow \mathit{coin\_spmf};$\\
$\qquad let \; m_0 = \neg u;$ & & $\qquad let \; m_0 = \neg u;$\\
$\qquad let \; m_1 = a \oplus (\neg u);$ && $\qquad let \; m_1 = a \oplus (\neg u);$ \\
$\qquad r \leftarrow R_1^{OT} \; m_0 \; m_1 \; b;$  &&  $\qquad r \leftarrow S_1^{OT} \; m_0 \; m_1;$\\
$\qquad \mathit{return_{spmf}} (u, r) \}$  && $\qquad \mathit{return_{spmf}} (u, r) \}$ 
\end{tabular} \caption{Simulator and real view of party A} \label{c}
\end{center}
\end{figure}

We now define a protocol (Fig \ref{b}) to compute an AND gate. The protocol uses OT as a black box to transfer $m_b$. Each party outputs an additive share of the desired AND gate output. This protocol is proved secure using the simulation-based approach. We use Lemma \ref{lem_OT} to prove security of this protocol in the semi-honest model. 
The real view and the simulator for party A are given in Fig \ref{c}. The simulator for party B, $S_B$, is constructed in an analogous way. Using these simulators we are able to show the AND gate protocol in Fig \ref{b} is information theoretically secure. 
\begin{lemma} Information theoretic security for the AND gate protocol is shown by the equalities
	$$R_A \; a \; b = S_A \; a \; \text{and} \; R_B \; a \; b = S_B \; b.$$
\end{lemma} 
 We have shown how a simple OT that uses a trusted initialiser can help to securely compute an AND gate. In general a trusted initialiser would not be necessary as one can use the N-P OT in the AND gate protocol. There is one technical issue with doing this. In the N-P OT we work with a group with multiplication but the AND gate protocol requires addition. In practice this is overcome by implementing the N-P OT using a ring (which has both operations), for which the DDH assumption holds. The proof would follow as in the proof given above, but an extension of the theory of rings in Isabelle is required for this - something we plan to develop in future work. 
 
\section{Conclusion}\label{sec:concs}

We have shown a general approach for capturing simulation-based cryptographic proofs in the computational model, building on Lochbihler's CryptHOL framework, and giving a
proof of the Naor-Pinkas OT protocol.  
We also have shown how out technique can be used to formally prove security of a simple two party protocol for an AND gate based on OT.

\paragraph{Future Work.} The work presented here is only a starting
point for the development of theory and examples of
simulation-based proofs.  Oblivious
Transfer is a fundamental cryptographic primitive which can be used to
construct generic protocols for MPC.
For example, Yao's garbled circuits use OT as a sub-protocol
to exchange garbled inputs, while the GMW protocol relies on OT
for computing AND gates.
Section~\ref{AND} took a first step
towards a formal proof of the GMW protocol.
Section \ref{AND} took a first step towards a formal proof of the GMW protocol. We plan to extend this work towards  formalising general MPC protocols.

\paragraph{Related Work.} Many formal techniques and tools have been devised which use the  symbolic model.  Work on formalising proofs in the computational model has begun more recently and is more challenging, requiring mathematical reasoning about probabilities and polynomial functions, besides logic.  
The CertiCrypt~\cite{CertiCrypt} tool built in Coq helped to capture the reasoning principles that were implemented directly in the dedicated interactive EasyCrypt tool~\cite{EasyCrypt}.  
Again in Coq, the Fundamental Cryptographic Framework~\cite{FCF}  
provides a definitional language for probabilistic programs, a theory that is used to reason about programs, and a library of tactics for game-based proofs.
Interactive tools seem invaluable for complex protocols or exploring new techniques, but  automatic tools are more practical when things become routine.  
CryptoVerif \cite{CryptoVerif} is a tool with a high level of automation but its scope only stretches to secrecy and authentication in protocols.  AutoG\&P~\cite{AutoGandP} is another automated tool dedicated to security proofs for pairing-based cryptographic primitives.
So far, all of these tools have been used to perform game-based cryptographic proofs and not simulation-based proofs.

\paragraph{Acknowledgements.}
We are deeply grateful to Andreas Lochbihler for providing and continuing to
develop CryptHOL and for his kind help given with using it. Also we are thankful to the reviewers for their comments regarding the presentation of our work.

\bibliographystyle{plain}
\bibliography{paper}

\end{document}